\documentclass[twocolumn]{aastex62}

\def \alf{\texttt{alf}}

\graphicspath{{./}{figures/}}

\usepackage{url}
\makeatletter
\g@addto@macro{\UrlBreaks}{\UrlOrds}
\g@addto@macro{\UrlBreaks}{\do\a\do\b\do\c\do\d\do\e\do\f\do\g\do\h\do\i\do\j
  \do\k\do\l\do\m\do\n\do\o\do\p\do\q\do\r\do\s\do\t%
  \do\u\do\v\do\w\do\x\do\y\do\z\do\A\do\B\do\C\do\D%
  \do\E\do\F\do\G\do\H\do\I\do\J\do\K\do\L\do\M\do\N%
  \do\O\do\P\do\Q\do\R\do\S\do\T\do\U\do\V\do\W\do\X%
  \do\Y\do\Z}
\makeatother

\submitjournal{ApJ}

\shortauthors{CHOI ET AL.}
\shorttitle{Quiescent Galaxy Colors}

\begin{document}

\title{The Imprint of Element Abundance Patterns on Quiescent Galaxy SEDs}

\author{Jieun Choi}
\affiliation{Harvard--Smithsonian Center for Astrophysics, 60 Garden Street, Cambridge, MA 02138, USA}
\author{Charlie Conroy}
\affiliation{Harvard--Smithsonian Center for Astrophysics, 60 Garden Street, Cambridge, MA 02138, USA}
\author{Benjamin D. Johnson}
\affiliation{Harvard--Smithsonian Center for Astrophysics, 60 Garden Street, Cambridge, MA 02138, USA}

\begin{abstract}
Stellar population synthesis (SPS) models have long struggled to reproduce observed optical through near-IR (NIR) spectral energy distributions (SED) of massive quiescent galaxies. We revisit this issue using a novel approach that combines the diagnostic power of full-spectrum fitting with recently updated stellar spectral libraries. First, we perform full-spectrum fitting of continuum-normalized stacked SDSS spectra in bins of velocity dispersion to infer their stellar population properties, such as the elemental abundances and age. Next, we use the resulting best-fit parameters to compute $ugriz$ colors, which are then compared to observed colors of the same galaxies. With this approach we are able to predict the $ugriz$ SEDs of low and high mass galaxies at the $\lesssim0.03$ mag level in nearly all cases. We find that the full optical through NIR SEDs of quiescent galaxies can be reproduced only when the spectrum is fit with a flexibility that is able to capture the behavior of the entire optical absorption line spectrum. The models include variations in individual elemental abundances, nebular emission lines, and the presence of young stellar components. The successful prediction of the SED shape from continuum-normalized spectra implies that the continuum information is largely contained in the narrow absorption features. These results also imply that attempts to model broadband photometry of quiescent systems will suffer from potentially significant biases if the detailed abundance patterns are not taken into account. 
\end{abstract}


%---------------------------------------------------------%
%---------------------------------------------------------%
\section{Introduction}

For decades, the stellar population synthesis (SPS) technique has been used to study the stars in quiescent galaxies \citep[e.g.,][]{Tinsley1980, Renzini2006, Walcher2011, Conroy2013}, which are primarily old, metal-rich, and enhanced in $\alpha$-capture elements relative to Fe-peak elements \citep[e.g.,][]{Worthey1992, Trager2000, Thomas2005, Kelson2006, Graves2007, Schiavon2007, Johansson2012, Conroy2014, Choi2016}. The current generation of models have reached a high level of sophistication, and are capable of inferring the detailed abundances of individual elements \citep[e.g.,][]{Schiavon2007, Thomas2011, Worthey2014, Conroy2014} and even the initial mass function \citep[IMF; e.g.,][]{vanDokkum2010, Conroy2012, Spiniello2012, Ferreras2013, LaBarbera2013, MartinNavarro2015, Sarzi2018}. Despite these successes, there still remain several unresolved problems. One such long-standing puzzle is the ``spectral energy distribution (SED) problem''; solar-metallicity, old simple stellar population (SSP) models predict optical and near-IR (NIR) colors that are generally too red compared to observed colors \citep{Eisenstein2001}. Several ideas have been proposed---a frosting of young stars, old and metal-poor stars, non-universal IMF, and $\alpha$ elements \citep{Wake2006, Maraston2009, Conroy2010, Ricciardelli2012, Vazdekis2015}---but there is still no strong consensus.

In particular, studies focused on the effect of $\alpha$ elements on the broadband SEDs of quiescent galaxies have been rather scarce, which is surprising given what we know from their spectra: their constituent stars are strongly $\alpha$ enhanced. \cite{Vazdekis2015} was the first to revisit the SED problem with self-consistent $\alpha$-enhanced models constructed from the BaSTI \citep{Pietrinferni2004, Pietrinferni2006} isochrones and MILES empirical \citep{SanchezBlazquez2006} and \cite{Coelho2005, Coelho2007} theoretical spectral libraries. These authors concluded that both $\alpha$-enhancement and a bottom-heavy IMF are necessary to improve the match with observed $ugri$ colors.

The results presented in \cite{Vazdekis2015} were based on the assumption that the $\alpha$ elements vary in lockstep while C and N track Fe, but the abundance patterns in real galaxies are much more complex \cite[e.g.,][]{Conroy2014}. Furthermore, there is no consistency in the treatment of C and N in SPS models in the literature. Sometimes either (or both) element is assumed to track the ``standard'' $\alpha$ elements such as Mg \citep[e.g.,][]{Trager2000, Gallazzi2005}, and in other cases, it is assumed to track Fe instead \citep[e.g.,][]{Coelho2007, Vazdekis2015}.

In this work, we revisit the puzzle using a novel approach in which colors are self-consistently predicted from models that are fit to the absorption line spectra. Our approach is similar in spirit to \cite{Tojeiro2011}, where the authors fit stacked SDSS spectra to infer parameters such as star formation history then computed evolutionary tracks in color and magnitude.

%---------------------------------------------------------%
%---------------------------------------------------------%
\section{Models}
\label{section:models}

%---------------------------------------------------------%
\subsection{Absorption Line Fitter (\alf{})}

The \alf{} code \citep{Conroy2012, Conroy2014, Choi2014, Conroy2018} is a full-spectrum fitting tool that models the optical to NIR spectra of systems containing older ($>1~\rm Gyr$) stellar populations using Markov Chain Monte Carlo (\texttt{emcee}; \citealt{ForemanMackey2013}). Here we provide a brief overview and refer the reader to \cite{Conroy2012} and \cite{Conroy2018} for details and also \cite{Choi2014} and \cite{Conroy2017} for performance tests with mock data. 

We start with the SSP models constructed from the \texttt{MIST} isochrones \citep{Dotter2016, Choi2016} and MILES+IRTF spectral libraries \citep{Villaume2017}. \alf{} makes use of age- and metallicity-dependent response functions---the response of the theoretical stellar spectra to abundance variations---to model systems with abundances that differ from those of stars that comprise the spectral library \citep{Tripicco1995, Korn2005, Lee2009, Sansom2013, Conroy2014}. We adopt \cite{Asplund2009} solar abundances as the reference values and vary the individual abundances at constant [Fe/H].

In its ``full'' mode, \alf{} simultaneously models the abundances of 19 elements and 20 other parameters, including $z$, $\sigma$, stellar age (old and young components), IMF slopes, and emission lines.  Among the 39 total parameters, several elements such as K, Sr, Ba, Eu, and Cu and parameters such as telluric absorption are of minor importance and thus inconsequential to the fitting process. \alf{} can also be operated in a limited, ``simple" mode to isolate the impact of elemental abundances, wherein the abundances of 9 elements (Fe, C, N, O, Na, Mg, Si, Ca, Ti), a single age, $z$, and $\sigma$ are retrieved. While \alf{}, by design, allows individual abundances to vary, it can also be modified to test assumptions such as lockstep variation in the $\alpha$ elements. In the latter case, we group Na, C, and N with the $\alpha$ elements because quiescent galaxies are also strongly enhanced in these elements \citep[e.g.,][]{Conroy2014}.

The \alf{} SSP models do not extend to the SDSS $u$ band ($\approx 3100\textrm{--}4000$~\AA) as the blue cutoff in the MILES library \citep{SanchezBlazquez2006} is at $3590$~\AA. However, access to the $u$ band is desirable since it is sensitive to abundance and age effects \citep{Coelho2007, Schiavon2007, Sansom2013, Vazdekis2015}. We correct for missing flux using the \texttt{Flexible Stellar Population Synthesis} models \citep[\texttt{FSPS};][]{Conroy2009, Conroy2010} generated with \texttt{ATLAS12/SYNTHE} synthetic stellar spectra \cite{Kurucz1970, Kurucz1993}, which extend down to 100~\AA{}. These solar-scaled models are first corrected for the abundance effects using the response functions and stitched to the \alf{} spectrum over $3600\textrm{--}3700$~\AA. Although the size of the correction is not negligible---approximately 50\% of the $u$ band flux falls below $3600$\AA{} for a 10~Gyr, $\rm [Fe/H]=0$ population---we find that the resulting $u$ magnitude is not strongly sensitive to the underlying FSPS model. For $\rm [Fe/H]=0$ stellar populations between 2 to 13~Gyr, $\Delta \rm [\alpha/Fe]=0.4$ results in only a $\lesssim 0.02$~mag difference in the resulting $u$ magnitude, which is well within the zeropoint uncertainties.

At the time of writing, \texttt{MIST} isochrones, which form the basis of our template SSPs, are unavailable for non-solar-scaled composition. Although a fully self-consistent model requires that the isochrones and spectral libraries have the same abundances (see e.g., Figure~9 in \citealt{Vazdekis2015}, which illustrates the small changes to the flux of the SSP spectra as a result of varying $\alpha$ abundances in the underlying isochrones), abundance variations influence the stellar spectra much more strongly than the isochrones, especially at old ages \citep{Coelho2007, Lee2009, Vazdekis2015}. Thus we proceed with SSPs constructed with solar-scaled isochrones.

%---------------------------------------------------------%
\subsection{The Effects of $\alpha$ Elements, C, and N on $ugriz$ Colors}
\label{section:cn}

\begin{figure}[!t]
\center
\includegraphics[width=0.45\textwidth]{{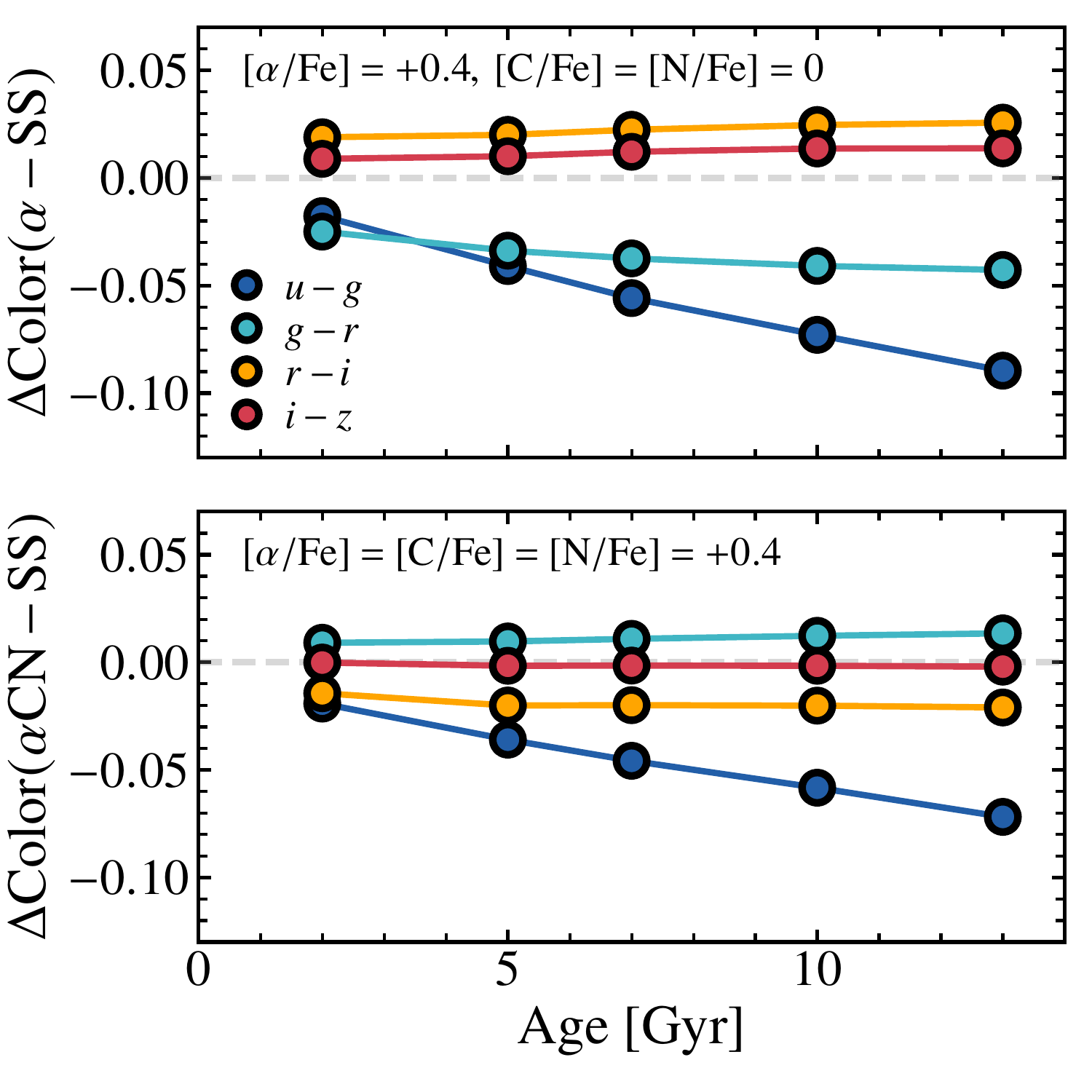}}
\caption{The effect of $\alpha$, C, and N abundances on $ugriz$ colors as a function of stellar population age. The colors are shown relative to those predicted for a solar-scaled model. The top panel shows the color predictions for a model with elevated $\alpha$ but solar-scaled C and N abundances, while the bottom panel shows colors for a model with enhanced $\alpha$, C, and N abundances. Overall, the $u-g$ color becomes bluer when the $\alpha$ abundance is increased, but the effect is reduced when C and N are increased alongside the $\alpha$ elements. $g-r$ and $r-i$ in these two panels show reversed trends depending on the treatment of C and N.}
\label{fig:alf_color_vs_age}
\end{figure}

In this section we utilize \alf{}-generated models to explore the impact of key groups of elements on the optical and NIR colors of quiescent galaxies. Note all magnitudes are quoted in the AB system \citep{Oke1983}. We first compute two sets of spectra at different ages, one representing the ``traditional'' $\alpha$-enhanced models in which $\rm [C/Fe] = [N/Fe] = 0$, and another representing models in which $\rm [\alpha/Fe] = [C/Fe] = [N/Fe] = +0.4$. In Figure~\ref{fig:alf_color_vs_age} we plot, as a function of age, $ugriz$ colors predicted from these models relative to those predicted from the solar-scaled models. Compared to solar-scaled models, standard $\alpha$-only enhancement leads to much bluer $u-g$ and $g-r$ and moderately redder $r-i$ and $i-z$ colors. This is due to increased flux at blue wavelengths \citep[e.g.,][]{Sansom2013, Vazdekis2015}. However, C and N enhancement leads to a smaller change in $u-g$, reversed trends for $g-r$ and $r-i$ (now redder and bluer, respectively), and $i-z$ that is indistinguishable from the solar-scaled case. This is due to strong CN absorption near $\sim3850$~\AA{} and $\sim4150$~\AA{} and the CH G-band at $4300$~\AA, which partially cancel the effect of the $\alpha$ enhancement. In summary, predicted colors of quiescent galaxies depend sensitively on the treatment of C and N as well as $\alpha$ elements.

\begin{figure*}[!t]
\center
\includegraphics[width=0.85\textwidth]{{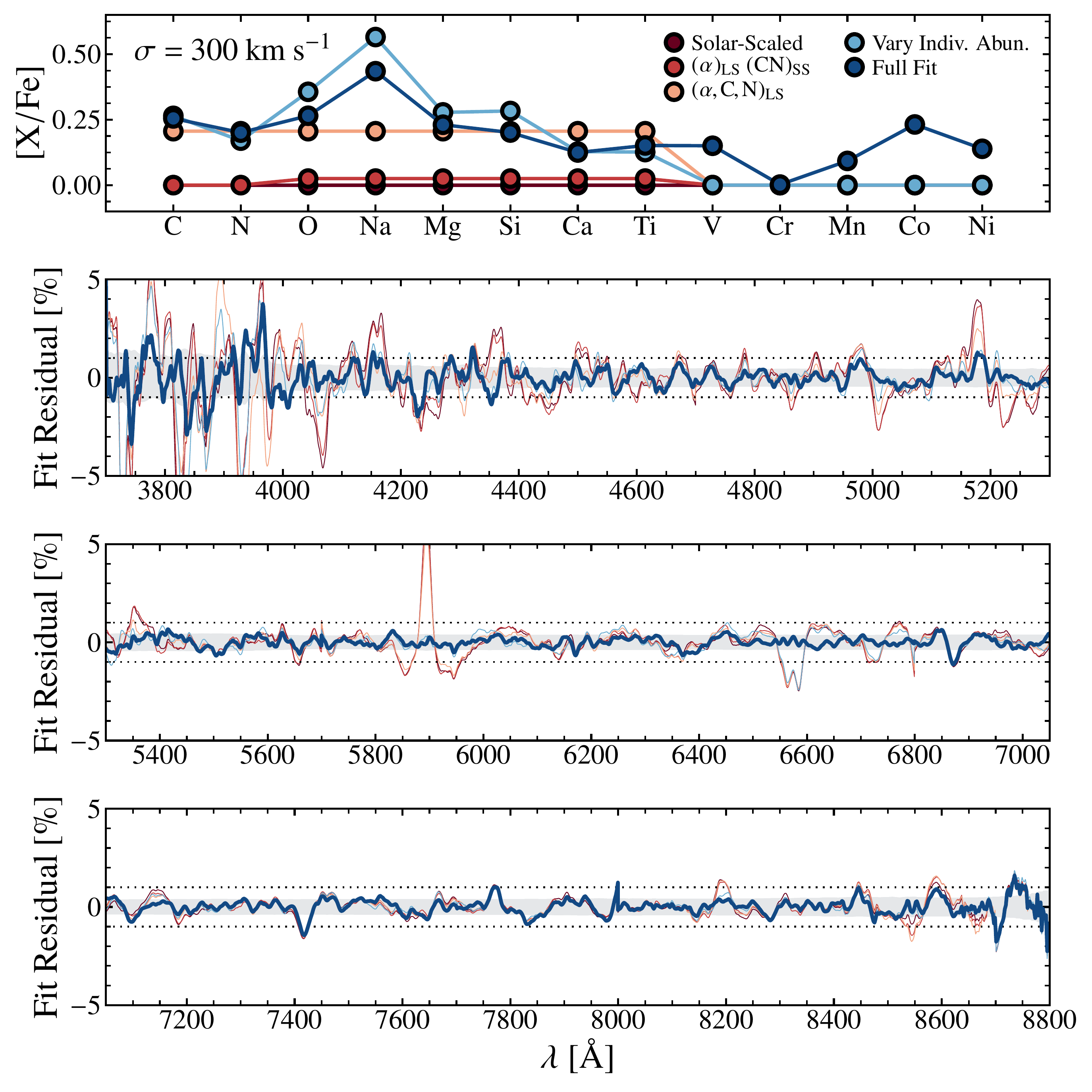}}
\caption{Summary of the results from fitting a stacked and continuum-normalized SDSS spectrum with \alf. Only the highest velocity dispersion bin is shown, as it harbors the largest deviations from solar-scaled abundances and therefore provides the strongest contrast between different modeling assumptions. The top panel presents the best-fit elemental abundances. The [Fe/H] values measured for this spectrum range from $-0.01$~dex to 0.21~dex ($\sigma_{\rm stat} \approx 0.02$~dex) depending on the fitting mode. The colors refer to five fitting modes: from navy to maroon, they are solar-scaled (SS); lockstep (LS) variation in $\alpha$ + Na and C + N tracking Fe; lockstep variation in $\alpha$, Na, C, and N; individual $\alpha$, Na, C, and N abundances; and individual abundances of all elements shown here plus additional parameters, such as the IMF slopes and emission line strengths. The bottom panels show the fractional residuals in the best-fit spectra. The flux uncertainties are shown as the gray band. Overall, only the full-fit (navy) can reproduce the strong spectral features over the entire wavelength range within the quoted uncertainties.}
\label{fig:sdss_fit}
\end{figure*}

\begin{figure*}[!t]
\center
\includegraphics[width=0.65\textwidth]{{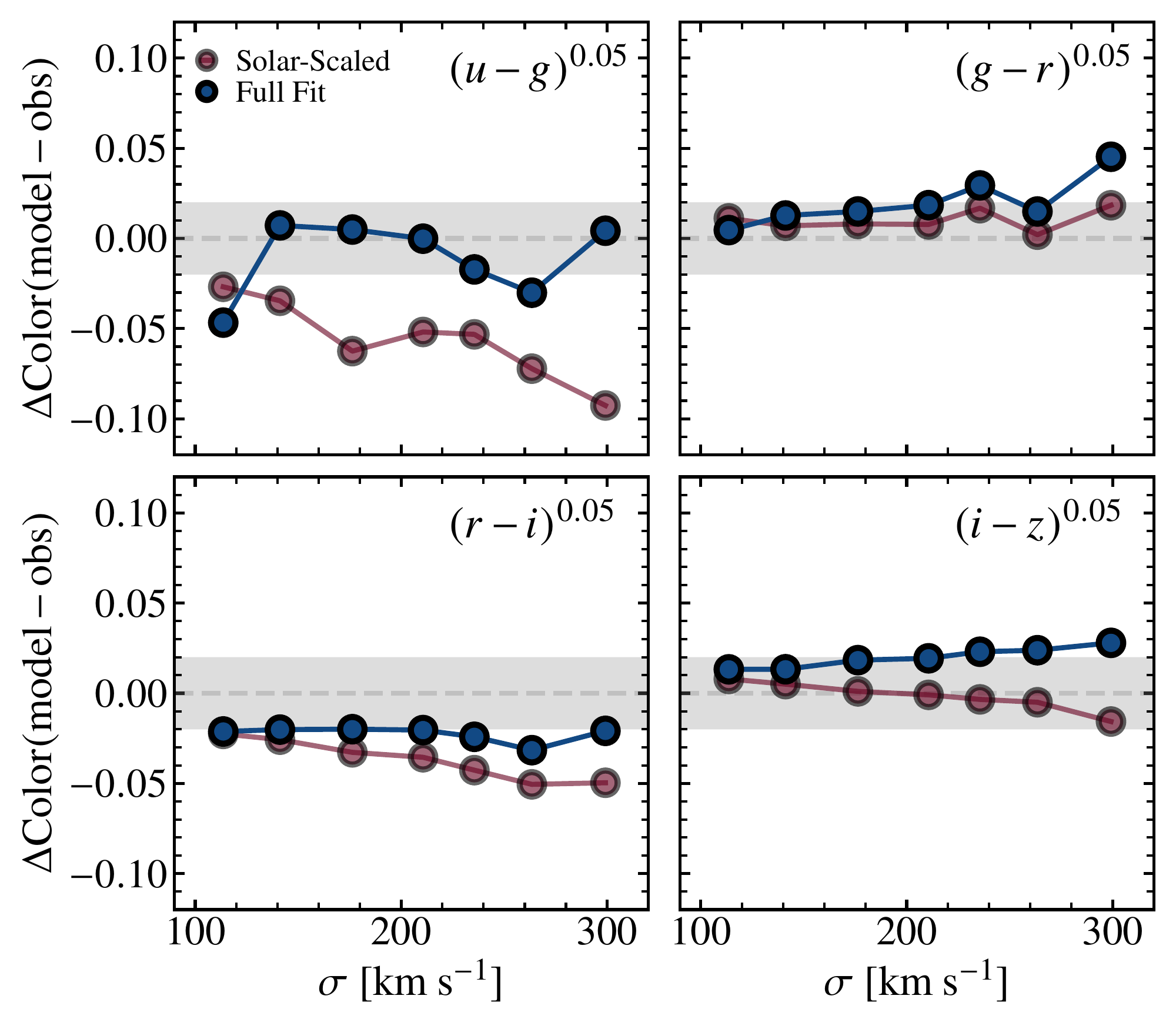}}
\caption{Comparison between predicted and observed colors as a function of $\sigma$. The predicted colors are the result of fitting the continuum-normalized, absorption line spectra and the observed colors correspond to the mean, extinction-corrected {\texttt{fiber mag}} in each $\sigma$ bin. The gray band represents the $0.02$~mag photometric zeropoint uncertainty.}
\label{fig:sdss_vs_alf_sigma_color}
\end{figure*}

\begin{figure*}[!t]
\center
\includegraphics[width=\textwidth]{{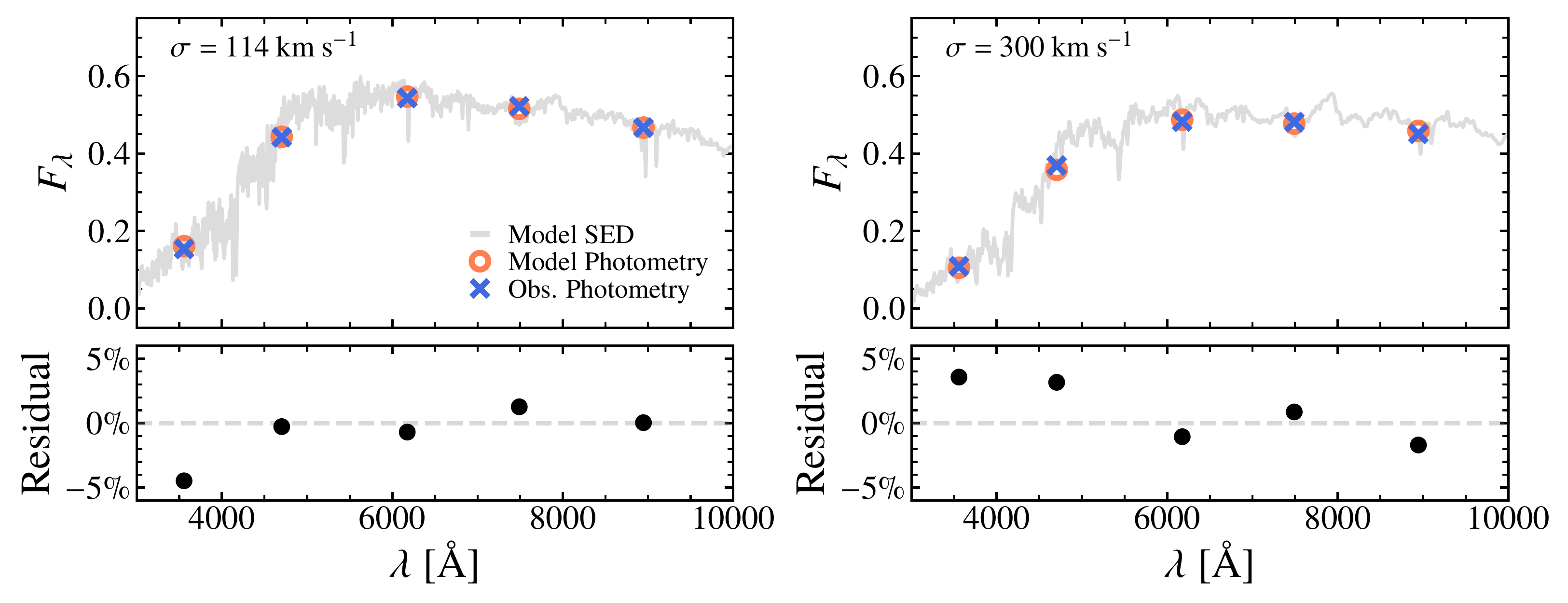}}
\caption{Predicted and observed $ugriz$ photometry shown alongside the full-fit SED for the lowest and highest $\sigma$ bins. The model photometry was computed from the model spectrum, shown in gray. The bottom panels show the fractional residuals.}
\label{fig:sed_comparison_flambda}
\end{figure*}

%---------------------------------------------------------%
%---------------------------------------------------------%
\vspace{2cm}
\section{Results}
\label{section:sdss_alf}

%---------------------------------------------------------%
\subsection{Fitting the Spectra}

The data in this work come from the SDSS Main Galaxy Survey \citep{Strauss2002} Data Release 7 \citep{Abazajian2009}. From the main sample, we select galaxies with $0.04\leq z \leq 0.06$, $E(B-V)<0.03$~mag, and no detected H$\alpha$ nor [OII]3727 emission. This subsample is further divided to seven velocity dispersion bins, ranging from $\log(\sigma/[\rm km\;s^{-1}]) = 2.2$ to 2.45, where the median redshift in each bin is $z\approx0.05$. For each bin, individual spectra are smoothed to the highest $\sigma$ and continuum-normalized by an $n=8$ polynomial prior to stacking. The resulting median S/N for each stack ranges from 240~\AA$^{-1}$ to 1203~\AA$^{-1}$.

The stacked spectra are fit with \alf{} over $3700\text{--}8850$~\AA{} in five modes:
\begin{enumerate}

\item Solar-Scaled\footnote{Technically, these are unmodified empirical SSPs, which means they reflect the abundance patterns of the underlying stellar spectral library. At near-solar metallicities, the abundance pattern is very close to solar-scaled.}: spectra are fit assuming solar-scaled abundances, wherein the template SSPs are left unmodified by response functions. This is the approach typically adopted in many current-generation SSP models, including \texttt{Starburst99} \citep{Leitherer1999}, \texttt{BC03} \citep{Bruzual2003}, \cite{Maraston2005}, and \texttt{FSPS} \citep{Conroy2009, Conroy2010}. 

\item {\tt $\rm (\alpha)_{LS}(CN)_{SS}$}: the $\alpha$ elements---O, Mg, Si, Ca, and Ti---and Na are fit in lockstep while the rest of the elements including C, N, and Fe-peak elements (e.g., Co, Ni) scale with Fe. This is the approach adopted in \citealt{Vazdekis2015}. 

\item {\tt $\rm (\alpha,C,N)_{LS}$}: $\alpha$, Na, C, and N are fit in lockstep and the Fe-peak elements scale with Fe. 

\item Vary Indiv.$\;$Abun.: $\alpha$, Na, C, and N abundances are fit individually and the Fe-peak elements scale with Fe.

\item Full Fit: all 39 parameters---additional SPS parameters such as emission lines and young and old age components plus the individual abundances of 19 elements, including the Fe-peak elements---are fit individually. 

\end{enumerate}

In options $1-4$ a single-burst star formation history (SFH) is adopted (i.e., SSP), while in option 5 the SFH consists of two age components.

Figure~\ref{fig:sdss_fit} shows results of fitting in these five options for the highest $\sigma$ bin, which represents the extreme case in terms of the stellar population properties. The top panel showing the abundance measurements clearly demonstrates that $\alpha$ elements generally do not vary in lockstep and that C, N, and Na are strongly enhanced in the most massive galaxies \citep[e.g.,][]{Johansson2012, Conroy2014}. We focus primarily on the elemental abundances here and revisit the other parameters in Section~\ref{section:discuss}. The bottom panels show fractional residuals for the best-fit spectra. Overall, the models are good fits to the stacked spectra, with root mean square residuals of around $1\textrm{--}5\%$ depending on the fitting mode. For the most part, only the full fit (navy) can reproduce the strong spectral features such as NaI near 5900~\AA{} and MgI near 5175~\AA. 

\begin{figure*}[!t]
\center
\includegraphics[width=0.85\textwidth]{{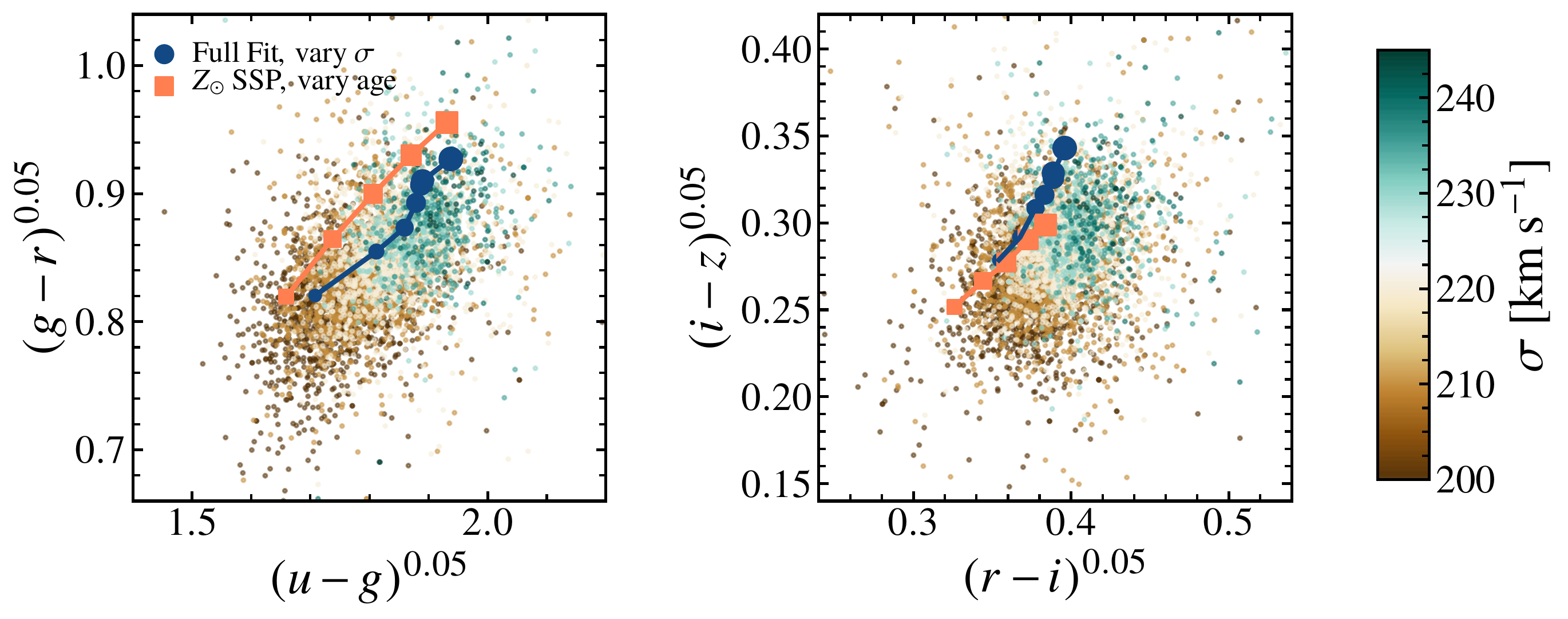}}
\caption{Observed $ugriz$ colors of individual galaxies. The navy points correspond to the colors predicted from the full fits to the SDSS stacked spectra and the orange points are colors computed for a solar metallicity SSP (5, 7, 9, 11, and 13~Gyr). Increasing point size corresponds to increasing $\sigma$ and age.}
\label{fig:sdss_vs_alf_color_color}
\end{figure*}

%---------------------------------------------------------%
\subsection{Predicted Colors from the Best-Fit Spectra}

The main result of this work is presented in Figure~\ref{fig:sdss_vs_alf_sigma_color}, where we compare the colors predicted from fitting the SDSS stacked and continuum-normalized spectra to the observed $ugriz$ \texttt{fiber mag} colors as a function of $\sigma$. Comparison with \texttt{fiber mag}---the integrated flux within the aperture of the $3^{\prime\prime}$ spectroscopic fiber---is ideal since it ensures that the observed light is coming from the same region within the galaxy for both the spectra and photometry. Note that \texttt{fiber mag} includes the effects of $2^{\prime\prime}$ seeing. We apply 0.04~mag and 0.02~mag zeropoint offsets\footnote{\url{ https://www.sdss.org/dr14/algorithms/fluxcal/\#FluxCalibration}} to the observed SDSS $u$ and $z$, respectively, to bring them to the AB system \citep{Oke1983}. We plot a gray band representing the likely zeropoint uncertainty of $0.02$~mag.

For simplicity, we only show the results for the full fit---all 19 elemental abundances are individually measured along with the IMF, two-component ages, etc.---in navy and the simple fit---only the total metallicity and a single age are measured---in faint red. Overall, the model and observed colors are in excellent agreement, which is especially remarkable given that \alf{} performs full-spectrum fitting on normalized spectra, suggests that the actual shape of the SED is encoded in the detailed absorption features. The comparison between the observed and predicted SEDs is presented in Figure~\ref{fig:sed_comparison_flambda}.

In Figure~\ref{fig:sdss_vs_alf_sigma_color}, the details of the model fit do not appear to have a significant impact on the $griz$ colors; both models accurately predict the colors to within $\lesssim 0.05$~mag. On the other hand, the simple model predicts $u-g$ that is too blue by $\sim 0.1$~mag for the highest $\sigma$ bin. This is because the best-fit age is driven to lower values ($\sim 7$~Gyr) to match the absorption features, which results in a much bluer SED. Note that $u-g$ encompasses the 4000~\AA-break and strong Ca features, which are sensitive to the age of the stellar population. Similarly, the mode in which the individual $\alpha$ abundances are allowed to vary but without additional parameters (light blue in Figure~\ref{fig:sdss_fit}) predicts $u-g$ that is too blue due to its preference for a young age. 

We performed two additional tests to investigate the importance of other model components. First, we refit the spectra in full mode with a fixed Kroupa IMF. The resulting colors did not change for the most part except for a small ($\approx 0.02$~mag) decrease in $i-z$ toward high $\sigma$ which, interestingly, improved the agreement with data. Second, we refit the spectra in full mode and forced all abundances to be solar-scaled to simulate the simple solar-scaled mode but with the complexities of the full mode. The resulting changes to the $griz$ colors were minuscule relative to the fiducial full mode, but the predicted $u-g$ became too red by $\approx 0.05$~mag in the two highest $\sigma$ bins. Taken together, these results suggest that the flexibility afforded by the full fit---abundance variations and the secondary young population---are necessary to simultaneously reproduce the spectrum and the SED.

Many previous studies that have compared model and observed colors have focused on color-color plots \citep[e.g.,][]{Conroy2010, Vazdekis2015}. In Figure~\ref{fig:sdss_vs_alf_color_color}, we present such comparison, but with the individual galaxies color-coded by their $\sigma$. The navy points correspond to predictions from the fiducial full fits to the SDSS spectra (same as in Figure~\ref{fig:sdss_vs_alf_sigma_color}) and the orange points are colors computed for a solar-scaled, solar metallicity SSP---representative of the model originally investigated in e.g., \citet[][]{Eisenstein2001}---for 5, 7, 9, 11, and 13~Gyr. Note the differences in scale along each color axis. These panels clearly illustrate the underlying trend in color with $\sigma$, highlighting the importance of carrying out model-data comparisons for the same subpopulation, e.g., in $\sigma$. But also note that comparisons in this color-color space are difficult to interpret owing to the significant scatter at fixed color. In contrast, the comparison in Figure~\ref{fig:sdss_vs_alf_sigma_color} offers a cleaner test of the models.

%---------------------------------------------------------%
%---------------------------------------------------------%
\section{Revisiting the ``SED problem''}
\label{section:discuss}

The now well-known discrepancy was first pointed out by \cite{Eisenstein2001}, who found that $g-r$ colors in the observed-frame (roughly rest-frame $u-g$) predicted by a single-burst, solar metallicity SPS model were too red by $\approx 0.08$~mag for the sample of $z<0.4$ SDSS luminous red galaxies (LRGs). The authors could not resolve this issue with more complex SFHs nor alternate SPS models, and these results were confirmed by \cite{Wake2006} who used additional data to extend the comparison to $0.45<z<0.8$.

\cite{Maraston2009} constructed SPS models using a new empirical library and found that $g-r$ and $r-i$ colors for $0.1<z<0.7$ galaxies could be correctly predicted with the addition of a sprinkling of old, metal-poor stars. Interestingly, the authors estimated the effect of $\alpha$ enhancement on the colors using the \cite{Coelho2007} models (recall that there is no C and N enhancement in these models) and found that it makes stellar spectra significantly bluer than is necessary to bring the observed and model colors into agreement.

The discrepancy was again confirmed by \cite{Conroy2010}, who also found that all models tend to over- and underpredict the strengths of $D_n4000$ and H$\delta_A$ spectral indices, leading to the conclusion that a secondary population of young or metal-poor stars are required.

\cite{Ricciardelli2012} also found that single-burst models predict $u-g$ and $g-r$ colors that are too red and $r-i$ colors that are too blue compared to galaxies at $z\approx0.04$, but a more complex model with contributions from young and/or metal-poor cannot simultaneously match the $ugri$ colors. They concluded that $\alpha$ enhancement likely plays an important role. New $\alpha$-enhanced SPS models were presented in \cite{Vazdekis2015}, who reported that $[\rm \alpha/Fe]>0$ improves the agreement in the colors {\it and} spectral features, though a bottom-heavy IMF is also necessary to match the observed colors of $z\sim0.04$ galaxies.

In contrast to previous work, we first fit the spectra then we computed the colors self-consistently from the resulting best-fit models. We also studied the effects of C and N on the resulting colors, which are qualitatively distinct from that of $\alpha$ elements (see Figure~\ref{fig:alf_color_vs_age}). Furthermore, by choosing a low redshift, we ensured that the model samples nearly rest-frame $u$ through $z$.

The several model assumptions we tested produced a range of spectral fits, with strong preference for the models that account for variations in the individual elemental abundances (Figure~\ref{fig:sdss_fit}). We also found that the best full-fit model prefers a small contribution ($\sim 1\textrm{--}10\%$ by mass, depending on $\sigma$) from a secondary young ($\sim 1\textrm{--}3$~Gyr) population, which is comparable to values that have been previously explored \citep[e.g.,][]{Ricciardelli2012}. Note that the effect of a young secondary population is similar though not identical to that of an old, metal-poor population. While the latter was not investigated in this work, previous work that has studied this effect \citep{Maraston2009} found that an old, metal-poor population contributing $\sim 3\%$ by mass noticeably alleviates the color discrepancies. Furthermore, we found that the full-fit mode recovers IMFs that are moderately bottom-heavy compared to a canonical Milky Way IMF \citep{Kroupa2001}, consistent with conclusions from \cite{Vazdekis2015}. A direct comparison with \cite{Vazdekis2015} is not possible since they adopted a unimodal IMF, but our best full-fit models yield mass-to-light-ratios ($M/L$) that are between one and two times the canonical $M/L_{\rm MW}$ (see also \citealt{Cappellari2012, Dutton2012, Conroy2012}).

Overall, we found that predicted $griz$ colors were nearly insensitive to the various assumptions, whereas $u-g$ showed as much as $\approx 0.15$~mag spread for the highest $\sigma$ bin, which harbors the largest departure from solar-scaled abundances. As demonstrated in Figure~\ref{fig:sdss_vs_alf_sigma_color}, $ugriz$ colors can be predicted to within $\lesssim 0.05$~mag for all $\sigma$ when we allow for full variation in the individual elemental abundances and the presence of a secondary, young population. 

There still remain small systematic discrepancies at the $\approx0.03$ mag level, whose origin is not yet clear and warrants further study. Future improvements to this work include extending the analysis to additional aperture-matched photometry, such as GALEX $FUV/NUV$ and 2MASS or UKIDSS $JHK$ data. Furthermore, as noted earlier, it will be important to investigate the effect of a metallicity distribution \citep[e.g.,][]{Maraston2009, Ricciardelli2012, Vazdekis2015}. Finally, although this work utilized solar-scaled isochrones given their minor effects on the broadband colors, it will be illuminating to test this assumption with new isochrones that reflect $\alpha$-enhanced or tailored elemental abundances.

\section{Summary}
\label{section:summary}
In this work we revisited the long-standing mismatch between observed and model colors of quiescent galaxies.  We adopted a novel approach in which the constraints from modeling the continuum-normalized spectrum spanning the wavelength range $\rm 3700\textrm{--}8850~\AA{}$ are used to self-consistently predict $ugriz$ colors of low-redshift quiescent galaxies. We summarize the two key conclusions:

\begin{itemize}
\item Accounting for C and N enhancement in addition to standard $\alpha$ element enhancement leads to a significantly different behavior of the model $ugriz$ colors compared to $\alpha$ enhancement alone. Given their impact on the spectra and their elevated abundance in massive quiescent galaxies, C and N should be treated individually rather than grouped with the rest of the $\alpha$ elements.

\item A sufficiently flexible model that allows for individual abundance variations as well as the presence of a secondary young population is able to simultaneously match the continuum-normalized absorption line spectra and the overall shape of the $ugriz$ SED for both low and high mass galaxies.  The spectra are reproduced to $\approx2$\% or better across the entire optical wavelength range, and the colors are reproduced to a level of $\lesssim0.05$ mag (in most cases to $\lesssim0.03$ mag).

\end{itemize}

We have demonstrated that the broadband colors of quiescent galaxies are sensitive to the detailed elemental abundance patterns at the level of $\approx0.05-0.1$ mag.  Broadband SED modeling of quiescent galaxies should therefore take these effects into account in order to minimize the possibility of biases and systematic uncertainties in derived parameters.

\acknowledgments 
We thank the anonymous referee for their constructive feedback, which improved the quality of this manuscript. CC acknowledges support from the Packard Foundation.

Funding for the SDSS and SDSS-II has been provided by the Alfred P. Sloan Foundation, the Participating Institutions, the National Science Foundation, the U.S. Department of Energy, the National Aeronautics and Space Administration, the Japanese Monbukagakusho, the Max Planck Society, and the Higher Education Funding Council for England. The SDSS Web Site is http://www.sdss.org/. 

The SDSS is managed by the Astrophysical Research Consortium for the Participating Institutions. The Participating Institutions are the American Museum of Natural History, Astrophysical Institute Potsdam, University of Basel, University of Cambridge, Case Western Reserve University, University of Chicago, Drexel University, Fermilab, the Institute for Advanced Study, the Japan Participation Group, Johns Hopkins University, the Joint Institute for Nuclear Astrophysics, the Kavli Institute for Particle Astrophysics and Cosmology, the Korean Scientist Group, the Chinese Academy of Sciences (LAMOST), Los Alamos National Laboratory, the Max-Planck-Institute for Astronomy (MPIA), the Max-Planck-Institute for Astrophysics (MPA), New Mexico State University, Ohio State University, University of Pittsburgh, University of Portsmouth, Princeton University, the United States Naval Observatory, and the University of Washington.

\end{document}